\documentclass[a4paper]{jpconf}
\usepackage[backend=biber,style=ieee]{biblatex}
\usepackage{graphicx}
\usepackage{xcolor}
\usepackage{hyperref}
\usepackage{indentfirst}
\usepackage[ruled,vlined]{algorithm2e}

\usepackage{caption}
\usepackage{subcaption}

\include{pythonlisting}
\usepackage{lineno}

\begin{document}
\title{Robust Neural Particle Identification Models}

\author{Aziz Temirkhanov$^{1}$, Artem Ryzhikov$^1$,  Denis Derkach$^{1}$, \\ Mikhail Hushchyn$^{1}$,  Nikita Kazeev$^{1}$  and Sergei Mokhnenko$^{1}$\newline on behalf of LHCb collaboration}

\address{$^1$ HSE University, 20 Myasnitskaya st., Moscow 101000, Russia}

\ead{atemirkhanov@hse.ru, aryzhikov@hse.ru}

\begin{abstract}
The volume of data processed by the Large Hadron Collider experiments demands sophisticated selection rules typically based on machine learning algorithms. One of the shortcomings of these approaches is their profound sensitivity to the biases in training samples. In the case of particle identification (PID), this might lead to degradation of the efficiency for some decays not present in the training dataset due to differences in input kinematic distributions. In this talk, we propose a method based on the Common Specific Decomposition that takes into account individual decays and possible misshapes in the training data by disentangling common and decay specific components of the input feature set. We show that the proposed approach reduces the rate of efficiency degradation for the PID algorithms for the decays reconstructed in the LHCb detector.
\end{abstract}

\section{Introduction}
\label{sec:intro}

Particle identification (PID) plays a crucial part in many high-energy physics analysis. A higher performance PID algorithm leads to a better background rejection and thus more precise results. In addition, the algorithm is required to work with high efficiency in the entire available wide range of signal topologies and kinematics and provide good discrimination for various analyses. Machine learning (ML) algorithms have gradually become the baseline approach for this task~\cite{ml_pid}. One large family of such algorithms is neural networks.

The PID algorithms of the LHCb experiment relies on several sub-detector systems~\cite{LHCb}. 
Compared to other LHC experiments, LHCb has more information for hadron identification, particularly the separation between charged hadrons, provided by the Ring Imaging Cherenkov (RICH) subdetector. 
The muon identification is provided by the muon chambers, while the responsibility of calorimeters is mainly leptons and photons. The LHCb experiment employs several machine learning solutions~\cite{ml_pid} to aggregate the information for the PID. These solutions proved to have overall high efficiency over a large volume of the phase space.
In this paper, we address one of the outstanding issues that can occur during the application of machine learning algorithms in the real-life scenario: algorithm's efficiency degradation in case the testing scenario is significantly different from training.

The paper is organized as follows. The problem formulation and description of current LHCb PID algorithms are given in Section~\ref{sec:problem}. Section~\ref{sec:csd} provides a description of the Common Specific Decomposition algorithm used for the robust PID models. Finally, the list of training and testing samples, and the results for the different PID models are presented in Section~\ref{sec:data}.

\section{Problem statement}
\label{sec:problem}

\begin{figure}[h]
    \centering
    \includegraphics[width=0.7\textwidth]{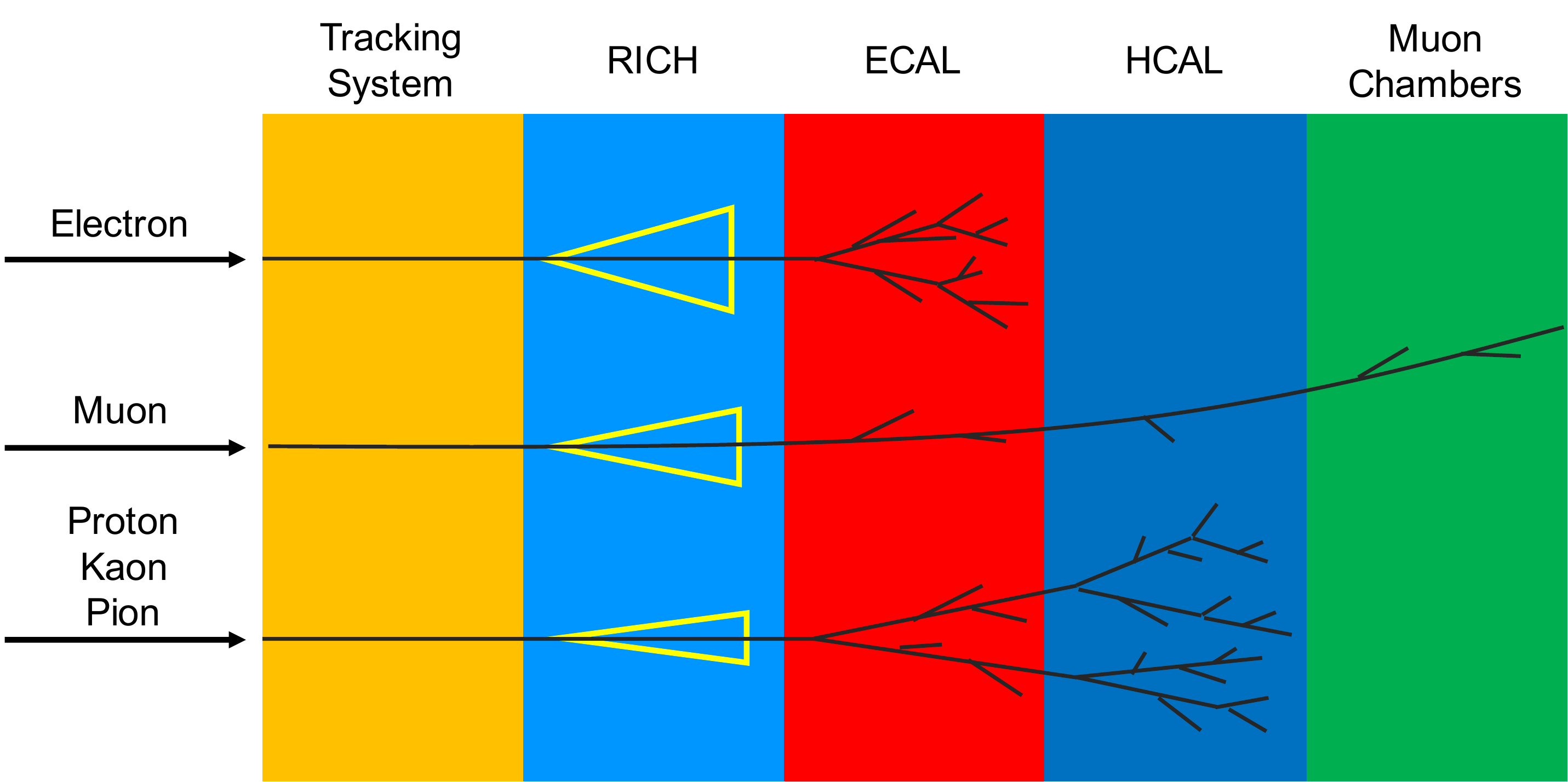}
    \caption{Illustration of different particle type responses in the LHCb systems. Figure from~\cite{2019EPJWC.21406011D}}
    \label{fig:pid_det}
\end{figure}

The LHCb detector is a single-arm forward spectrometer covering the pseudorapidity range $2 < \eta < 5$, described in detail in Ref.~\cite{LHCb}. Identification of various final state particles is performed by combining together the information from the LHCb subdetectors, namely from ring-imaging Cherenkov detectors~(RICH), the electromagnetic and hadronic calorimeters, muon chambers and tracking system, as demonstrated in Fig.~\ref{fig:pid_det}. Apart from the physics motivated likelihood observables based on observable subdetector responses~\cite{likelihood_subd}, track geometry variables and different detector flags are also used. In addition, the muon identification~\cite{muon_id} and calorimeter information about neutral clusters~\cite{neutral_clusters} are also used. 

The objective of the PID algorithm is to identify the charged particle type associated with a given track. In the LHCb experiment, there are five relevant particle species: electron, muon, pion, kaon, and proton. A sixth hypothesis corresponds to ghost tracks, which do not correspond to a real particle that passed through the detector. 

\begin{figure}[h]
    \centering
    \includegraphics[width=1\textwidth]{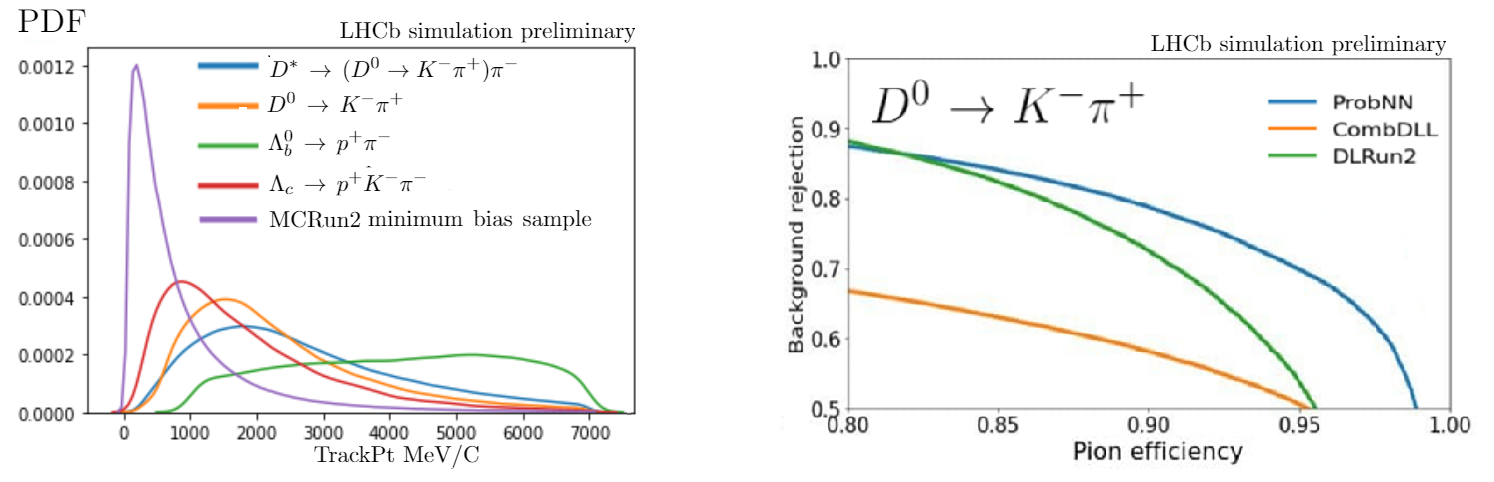}
    \caption{Distributions of transverse momentum for different decays in the simulated samples (left) and performance of the PID models on $D^{0} \to K^-\pi^+$ decay for the pion-vs-all task (right).}
    \label{fig:dist}
\end{figure}

The classification models for this task are trained on simulated events generated for the LHCb data taking condition for 2015-2018 data taking period with minimum bias trigger selection. In this study, we also use dedicated simulated samples of correctly reconstructed decays as a source of correctly reconstructed particles of a certain type. Particles  in these decays might have distribution different from the training sample as demonstrated in Fig.~\ref{fig:dist}. Such differences may cause a quality reduction of the classification models in this regions. The aim of this research is to make PID algorithms~\cite{ml_pid} robust to differences in the kinematic distributions between the events used for training and events the algorithm faces in real use. The research is focused on neural networks only.

In this work, we compare our PID algorithm with the following three models: \textbf{ProbNN}, \textbf{CombDLL}, and \textbf{DLRun2}. ProbNN is a neural network with a single hidden layer. CombDLL is a combination of differential log-likelihoods from the detector's subsytems. DLRun2 is a deep neural network with 3 wide hidden layers. Neural-based models are trained on minimum bias sample. The performance of the neural-based solutions is approximately the same when tested on the minimum bias sample.

We notice that quality of the algorithms might deteriorate in some regions. The performance of the classifiers is tested with a sample of  $D^{0} \to K^-\pi^+$ signal candidates (Fig~\ref{fig:dist}). As can be seen from the Figure, DLRun2 algorithm's quality is lower than its counterparts.



\section{CSD algorithm}
\label{sec:csd}

\begin{figure}[!htb]
    \centering
    \includegraphics[width=0.4\textwidth]{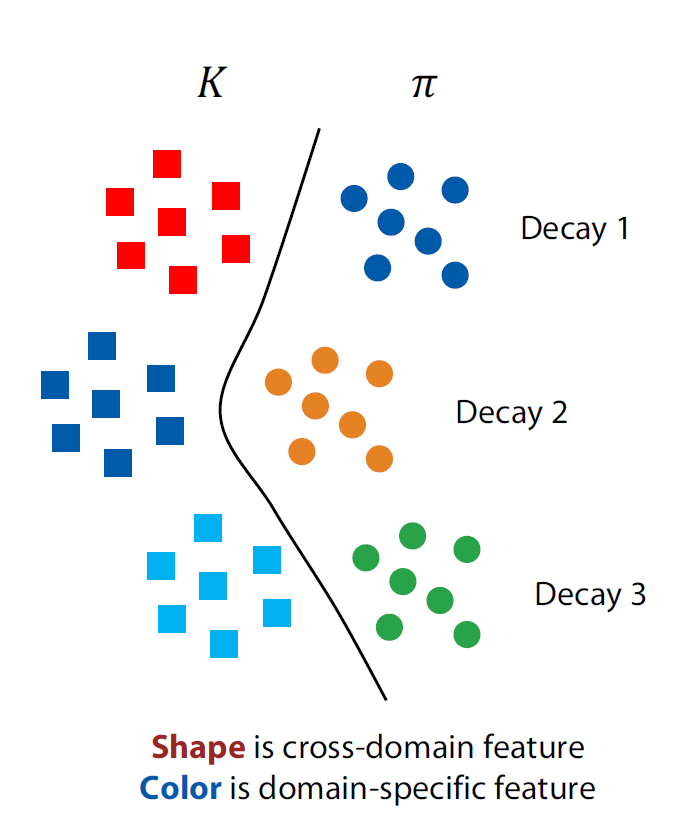}
    \caption{The CSD assumption.}
    \label{fig:assumption} 
\end{figure}

To address the quality degradation, we use domain adaptation techniques~\cite{Vihari_CSD}. 
The idea laying behind the method is as follows: assume that there exist common features, whose correlation with the target is preserved in all decays, and decay-specific features, whose correlation differs from decays to decay (Fig~\ref{fig:assumption}). A classifier that relies solely on common features is robust to domain shifts. 

To achieve that, CSD algorithm use a weighted combination of these three terms in training objective: orthonormality regularizes and two cross-entropy losses between labels and distributions computed from the common and domain-specific features. 

This approach allows to compute low-rank decomposition along with a training networks parameters in a single layer. Hence, we replace final classification layer of DLRun2 model by CSD layer~\cite{Vihari_CSD} as shown in Fig.~\ref{fig:assumption}.

\section{Data and results}
\label{sec:data}

For the experiment we use simulated minimum bias sample and pions from several decay channels: $D^{0} \to K^-\pi^+$, $D^* \to ({D^{0} \to K^-\pi^+})\pi^-$, $\Lambda_c \to p^+K^-\pi^-$, and $\Lambda_b^0 \to p^+\pi^-$. We train CSD model similarly to DLRun2 model on minimum bias sample. Then, we test every model on simulated run2 sample and each decay channel. For the CSD model an additional round of training with mixture of minimum bias sample and subset of listed above decays has been done in order to teach model to generalize in pion domain.

In addition, we design a nearest neighbors test in the following way. For a correctly reconstructed particle in the decay chain we choose the kinematically closest particle of minimum bias sample for the different species. We thus mimic the misreconstruction real-life scenario.

\begin{figure}[!htb]
    \centering
    \includegraphics[width=0.9\textwidth]{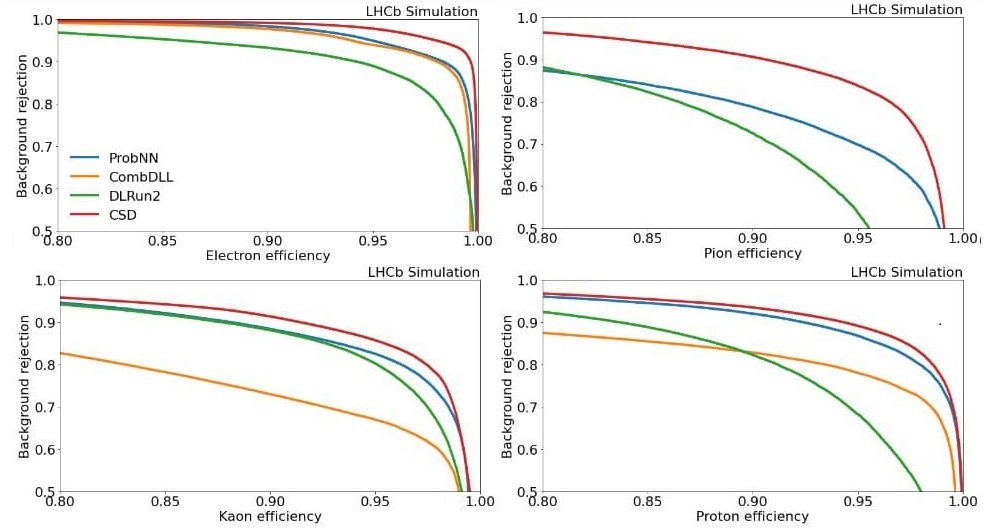}
    \caption{ROC AUCs for nearest neighbours test, the background sample is constructed of the best matching candidates as described in the text.}
    \label{fig:csd_rocs} 
\end{figure}

 Table~\ref{tab:mcrun2} contains ROC AOC (i.e. Area over ROC curve) increment with respect to DLRun2 (uncertainty does not exceed 0.0008 as obtained by a bootstrap procedure). The result for nearest neighbors test is shown in Fig.~\ref{fig:csd_rocs}.

\begin{table}[!htb]
\centering
\begin{tabular}[c]{|c| c | c | c | c | c | c|}
 \hline
 Model & Ghost & Electron & Muon & Pion & Kaon & Proton \\ [0.5ex] 
 \hline
 DLRun2 & 100\% & 100\%& 100\% & 100\% & 100\% & 100\% \\
 \hline
  ProbNN & $+41\% $ & $+14\% $ & $+418\% $ & $+51\% $ & $20\% $ & $+21\% $  \\
  \hline
  CombDLL & -- & $+71\%$ & $+389\%$ & $+979\%$ & $+89\% $ & $+98\% $ \\
  \hline
  \textbf{CSD} & $\textbf{-11\%}$ & $\textbf{-42\%} $ & $\textbf{-47\%}$ & $\textbf{-0.21\%}$ & $\textbf{-10\%}$ & $\textbf{-11\%}$ \\
  \hline
\end{tabular}
\caption{\label{tab:mcrun2}ROC AOCs change for different PID models for paricle-vs-all task. DLRun2 AOC is taken as 100\%.}
\end{table}

The quality on those 4 decays itself rises significantly, as it demonstrated in Table ~\ref{tab:pion}.

\begin{table}[!htb]
\centering
\begin{tabular}[c]{|c| c | c | c | c|}
 \hline
 Model & $D^* \to ({D^{0} \to K^-\pi^+})\pi^-$ & $\Lambda_c \to p^+K^-\pi^-$ & $\Lambda_b^0 \to p^+\pi^-$   & $D^{0} \to K^-\pi^+$ \\ [0.5ex] 
 \hline
 DLRun2 & 100\%  & 100\% & 100\% & 100\%  \\
 \hline
  \textbf{CSD} & $\textbf{-38\%}  $ & $\textbf{-71\%}  $ & $\textbf{-50\%}  $ & $\textbf{-15\%}  $  \\
  \hline
\end{tabular}
\caption{\label{tab:pion}ROC AOC for the pionic decays unseen by the CSD algorithm during training. DLRun2 AOC is taken as 100\%.}
\end{table}

In addition to previous tests, we cross-check that the obtained solution has some generalization power. In order to prove this,  we take two decays containing muons: $B^+ \to {K^+(J/\psi\to{\mu^+\mu^-})}$ and $B^+ \to {\mu^+\mu^-(K^*\to{K^+\pi_0)}}$ and compare CSD and DLRun2 quality. ROC AUC error rate for proposed method almost halved. Note, that models has never seen muon decays outside a minimum bias sample before.
Table ~\ref{tab:muon} illustrates that model can generalize well over different domains.

\begin{table}[!htb]
\centering
\begin{tabular}[c]{|c| c | c |}
 \hline
 Model & $B^+ \to {K^+(J/\psi\to{\mu^+\mu^-})}$ & $B^+ \to {\mu^+\mu^-(K^*\to{K^+\pi_0)}}$  \\ [0.5ex] 
 \hline
 DLRun2 & 100\%  & 100\% \\
 \hline
  \textbf{CSD} & $\textbf{-97\%}  $ & $\textbf{-98\%} $   \\
  \hline
\end{tabular}
\caption{\label{tab:muon}ROC AOC on muonic decays. DLRun2 AOC is taken as 100\%.}
\end{table}

\section{Conclusion}
\label{sec:conclusion}

The degradation of machine learning solutions quality for the PID problem in the specific parts of the phase space is addressed. 
The CSD algorithm based on the separation of common and domain-specific features has shown promising results.
The algorithm is able to select common features even for the decays that are not present in the original domains. The obtained algorithm shows higher stability with respect to previously presented thus giving substantial increase of solution's quality for particular case. The method thus can be used in many high-energy physics applications.

\section*{Acknowledgement}
The research leading to these results has received funding from Russian Science Foundation under grant agreement n° 17-72-20127. This research was supported in part through computational resources of HPC facilities at HSE University~\cite{HSE_HPC}. 


\nocite{python}
\nocite{pytorch}

\printbibliography

@Misc{python,
% author =   {{Python project}},
% title =    {{"Python" [software], version 3.6.7}},
% year = 2018,
% howpublished = {{Available from \url{https://www.python.org/downloads/release/python-367/} [accessed 2018-12-20]}}
% }

@Misc{pytorch,
% author = {{PyTorch project}},
% title = {{"PyTorch" [software], version 1.0.0}},
% year = 2018,
% howpublished = {{Available from \url{https://github.com/pytorch/pytorch} [accessed 2018-12-20]}}
% }

@Misc{ml_pid,
author = {{Derkach D {\itshape et al.}}},
year = 2018,
title = {{Machine-Learning-based global particle-identification algorithms at the LHCb experiment}},
howpublished = {{{\it J. Phys.: Conf. Ser.} \textbf{1085} 042038}}
}

@Misc{LHCb,
author = {{The LHCb Collaboration}},
year = 2008,
title = {{The LHCb Detector at the LHC}},
howpublished = {{{\it JINST} \textbf{3} S08005}}
}

@Misc{likelihood_subd,
author = {{The LHCb RICH group}},
year = 2013,
title = {{Performance of the LHCb RICH detector at the LHC}},
howpublished = {{{\it Eur. Phys. J. C} \textbf{73} 2431}}
}

@Misc{muon_id,
author = {{Archilli F {\itshape et al.}}},
year = 2013,
howpublished = {{\itshape JINST} \textbf{8} P10020 ({\itshape Preprint} 1306.0249)}
}

@Misc{neutral_clusters,
author = {{Deschamps O, Machefert F P, Schune M H, Pakhlova G and Belyaev I}},
year = 2003,
title = {{Photon and neutral pion reconstruction}},
howpublished = {Tech. Rep. LHCb-2003-091 CERN Geneva \url{https://cds.cern.ch/record/691634}}
}

@Misc{Vihari_CSD,
  title={Efficient Domain Generalization via Common-Specific Low-Rank Decomposition},
  author={{Vihari Piratla, Praneeth Netrapalli, Sunita Sarawag}},
  journal={ICML},
  year= 2020,
  howpublished =  "\url{https://arxiv.org/abs/2003.12815}"
  }

@article{HSE_HPC,
  title={{HPC Resources of the Higher School of Economics}},
  author={{Kostenetskiy}, P.S. and {Chulkevich}, R.A. and {Kozyrev}, V.I.},
  journal={Journal of Physics: Conference Series.},
  year=2021,
  volume = {1740},
  number = {1},
  pages = {012050},
  doi = {10.1088/1742-6596/1740/1/012050}
  }

@INPROCEEDINGS{2019EPJWC.21406011D,
       author = {{Derkach}, Denis and {Hushchyn}, Mikhail and {Kazeev}, Nikita},
        title = "{Machine Learning based Global Particle Identification Algorithms at the LHCb Experiment}",
    booktitle = {European Physical Journal Web of Conferences},
         year = 2019,
       series = {European Physical Journal Web of Conferences},
       volume = {214},
        month = jul,
          eid = {06011},
        pages = {06011},
          doi = {10.1051/epjconf/201921406011},
       adsurl = {https://ui.adsabs.harvard.edu/abs/2019EPJWC.21406011D}
}

\end{document}